\documentclass[prl,twocolumn,superscriptaddress]{revtex4-2}
\usepackage{amssymb}
\usepackage{amsfonts}
\usepackage{bbm}
\usepackage{mathrsfs}
\usepackage{graphics,graphicx,epsfig,bm,amsmath,amsthm,amssymb}
\usepackage{bm}
\usepackage{bbm}
\usepackage{longtable}
\usepackage{multirow}
\usepackage{array}
\usepackage{color}
\usepackage{cases}
\usepackage{booktabs }
\usepackage[usenames,dvipsnames]{xcolor}

\usepackage{xcolor}
\usepackage[none]{hyphenat}
\usepackage{float}
\usepackage{amsmath}
\usepackage[a4paper,colorlinks=true,
linkcolor=blue,citecolor=blue,
pdfauthor={ },
pdftitle={ },
pdfsubject={ },
pdfkeywords={ }]{hyperref}

\begin{document}

\title{Near-Quantum-limited Haloscope Detection of Dark Photon Dark Matter Enhanced by a High-Q Superconducting Cavity}

\author{Runqi Kang}
\affiliation{CAS Key Laboratory of Microscale Magnetic Resonance and School of Physical Sciences, University of Science and Technology of China, Hefei 230026, China}
\affiliation{CAS Center for Excellence in Quantum Information and Quantum Physics, University of Science and Technology of China, Hefei 230026, China}
\affiliation{Hefei National Laboratory, Hefei 230088, China}

\author{Man Jiao}
\affiliation{CAS Key Laboratory of Microscale Magnetic Resonance and School of Physical Sciences, University of Science and Technology of China, Hefei 230026, China}
\affiliation{Institute of Quantum Sensing and School of Physics, Zhejiang University, Hangzhou 310027, China}

\author{Yu Tong}
\affiliation{CAS Key Laboratory of Microscale Magnetic Resonance and School of Physical Sciences, University of Science and Technology of China, Hefei 230026, China}
\affiliation{CAS Center for Excellence in Quantum Information and Quantum Physics, University of Science and Technology of China, Hefei 230026, China}

\author{Yang Liu}
\affiliation{International Quantum Academy, Shenzhen 518048, China}

\author{Youpeng Zhong}
\affiliation{Shenzhen Institute for Quantum Science and Engineering, Southern University of Science and Technology, Shenzhen, China}
\affiliation{International Quantum Academy, Shenzhen 518048, China}

\author{Yi-Fu Cai}
\affiliation{Deep Space Exploration Laboratory/School of  Physical Sciences,
University of Science and Technology of China, Hefei, Anhui 230026, China}
\affiliation{CAS Key Laboratory for Researches in Galaxies  and
Cosmology/Department of Astronomy, School of Astronomy and Space Science,
University of Science and Technology of China, Hefei, Anhui 230026, China}

\author{Jingwei Zhou}
\email{zhoujw@ustc.edu.cn}
\affiliation{CAS Key Laboratory of Microscale Magnetic Resonance and School of Physical Sciences, University of Science and Technology of China, Hefei 230026, China}
\affiliation{CAS Center for Excellence in Quantum Information and Quantum Physics, University of Science and Technology of China, Hefei 230026, China}
\affiliation{Hefei National Laboratory, Hefei 230088, China}

\author{Xing Rong}
\email{xrong@ustc.edu.cn}
\affiliation{CAS Key Laboratory of Microscale Magnetic Resonance and School of Physical Sciences, University of Science and Technology of China, Hefei 230026, China}
\affiliation{CAS Center for Excellence in Quantum Information and Quantum Physics, University of Science and Technology of China, Hefei 230026, China}
\affiliation{Hefei National Laboratory, Hefei 230088, China}

\author{Jiangfeng Du}
\email{djf@ustc.edu.cn}
\affiliation{CAS Key Laboratory of Microscale Magnetic Resonance and School of Physical Sciences, University of Science and Technology of China, Hefei 230026, China}
\affiliation{CAS Center for Excellence in Quantum Information and Quantum Physics, University of Science and Technology of China, Hefei 230026, China}
\affiliation{Hefei National Laboratory, Hefei 230088, China}
\affiliation{Institute of Quantum Sensing and School of Physics, Zhejiang University, Hangzhou 310027, China}

\date{\today}

\begin{abstract}

We report new experimental results on the search for dark photons based on a near-quantum-limited haloscope equipped with a superconducting cavity.
The loaded quality factor of the superconducting cavity is $6\times10^{5}$,
  so that the expected signal from dark photon dark matter can be enhanced by more than one order compared to a copper cavity.  A Josephson parametric amplifier with a near-quantum-limited noise temperature has been utilized to minimize the noise during the search.
Furthermore, a digital acquisition card based on field programmable gate arrays has been utilized to maximize data collection efficiency with a duty cycle being 100$\%$.
This work has established the most stringent constraints on dark photons at around 26.965 $\mu$eV.
In the future, our apparatus can be extended to search for other dark matter candidates, such as axions and axion-like particles, and scrutinize new physics beyond the Standard Model.

\end{abstract}

\maketitle

\begin{figure}[t]
  \centering
  \includegraphics[width=1\columnwidth]{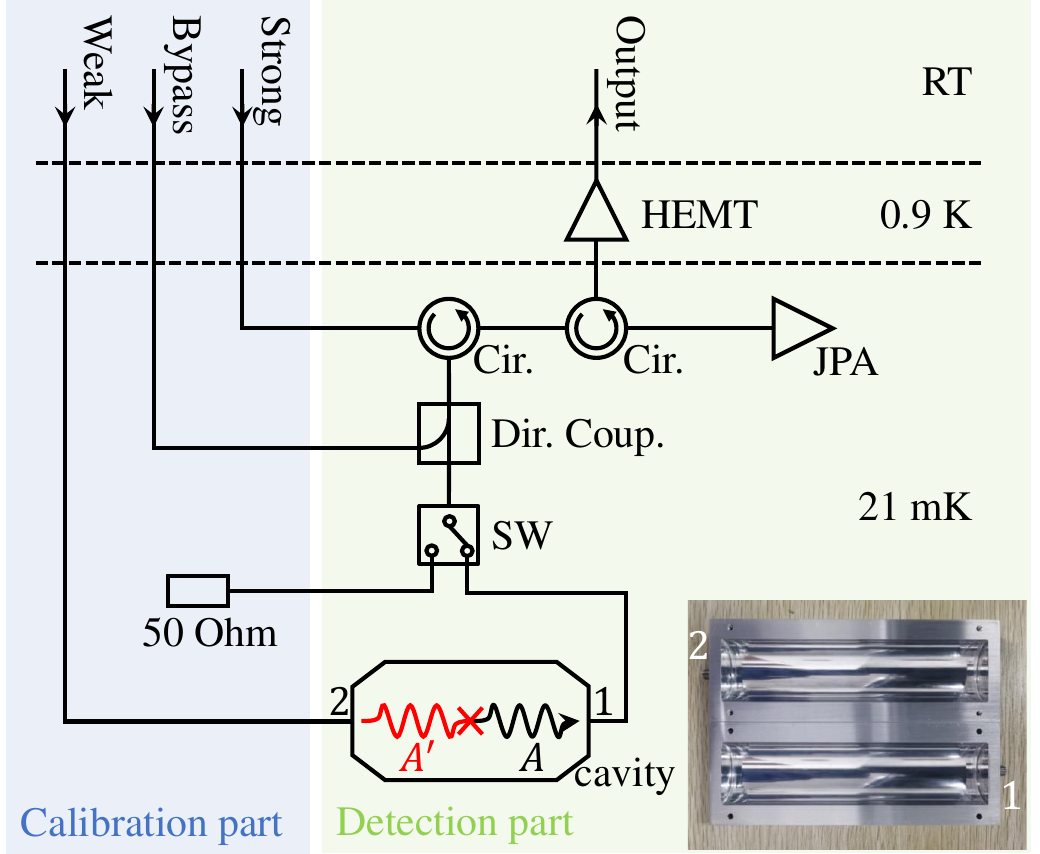}
  \caption
  {
    Simplified diagram of the experimental apparatus.
    The green box stands for the detection part.
    It includes a cylindrical superconducting cavity of high quality factor and an output line with an amplifier chain for the measurement of the signal.
    $A'$ and $A$ refer to the dark photon and the photon, respectively.
    The blue box stands for the calibration part.
    It includes three input lines and a 50 Ohm load.
    The inset is a photograph of the cavity.
    Its inner side is finely polished for a high Q factor.
    MW signals are coupled to the transmission lines through two coaxial probes.
    The two probes are fabricated on the opposite sides of the cavity to avoid their direct coupling.
  }
  \label{Figure1}
\end{figure}

Dark matter is one of the most enigmatic and salient topics in modern physics \cite{1906Poincare,1987Trimble,2010Feng,2018Bertone,2020Filippi}.
Although there are abundant phenomena that point to dark matter, its existence yet remains a mystery \cite{1933Zwicky,2003Refregier,2010Pospelov,2011Komatsu}.
The dark photon is an important candidate for dark matter, and also a possible portal that connects the Standard Model sector and the dark sector.
It is claimed to be helpful to explain many unsolved problems in physics, such as the velocity discrepancy of galaxies \cite{1933Zwicky}, the cosmic ray anomaly \cite{2008Chang,2009Cholis}, the muon anomalous magnetic moment \cite{2021Abi,2021Cazzaniga} and the W-boson mass anomaly \cite{2022Aaltonen,2022Thomas,2023Zhang}.

The dark photon is a spin-1 massive boson that rises from a slight extension of the Standard Model \cite{1984Galison,1986Holdom1,1986Holdom2}.
It couples to the ordinary photon via the following Lagrangian:
\begin{equation}
  \mathcal{L}  = - \frac{1}{4}F_{\mu\nu}^{2} - \frac{1}{4}V_{\mu\nu}^{2} - \frac{m_{A'}^{2}}{2}A'_{\mu}A'^{\mu} + \frac{\chi}{2}F_{\mu\nu}V^{\mu\nu},
\label{Lagrangian}
\end{equation}
where $A_{\mu}$ and $A'_{\mu}$ are the gauge fields of the ordinary photon and the dark photon, respectively,
$F_{\mu\nu} = \partial_{\mu}A_{\nu} - \partial_{\nu}A_{\mu}$ and $V_{\mu\nu} = \partial_{\mu}A'_{\nu} - \partial_{\nu}A'_{\mu}$ are the corresponding electromagnetic field tensors, $m_{A'}$ is the mass of the dark photon, and $\chi$ is the kinetic mixing, through which the dark photon interacts with the ordinary photon.
The dark photon mass, $m_{A'}$, is a free parameter that can take a wide range of values, while the kinetic mixing $\chi$ can be very small.
These two features render the search for the dark photon an extremely challenging task.

The haloscope is one of the most widely used methods for dark photon searching.
It is based on the fact that dark photons can convert to ordinary photons in a microwave cavity due to the kinetic mixing term \cite{1983Sikivie,1985Sikivie,2021Caputo}.
The power of the dark photon signal is
\begin{equation}
  P_{\rm s}(f)  = \frac{m_{A'}\rho_{A'}}{\hbar}\chi^{2}VC\frac{Q_{L}Q_{a}}{Q_{L}+Q_{a}}\frac{\beta}{1+\beta}L(f,f_{c},Q_{L}),
\label{Power}
\end{equation}
where $\rho_{A'} = 0.45$ GeV/cm$^{3}$ is the density of dark matter \cite{2014Read}, $\beta$ is the coupling coefficient between the cavity and the output port,
  $f_{c}$, $V$, $C$ are the resonant frequency, the volume and the filling factor of the cavity, respectively,
  $Q_{L}$ and $Q_{a}$ are the quality factor of the cavity and dark photons, respectively,
  and $L(f,f_{c},Q_{L}) = [1+4(Q_{L}(f-f_{c})/f_{c})^{2}]^{-1}$ is the Lorentzian line shape.
Many groups across the world have been working on searching for dark photons.
Projects such as MADMAX \cite{2020Gelmini}, QUAX \cite{2021Alesini, 2022Alesini} and ADMX \cite{2022Cervantes1} use cavities to conduct resonant detections.
 The SQuAD experiment \cite{2021Dixit} utilizes qubits to enhance the sensitivity.
 Experiments like FUNK \cite{2020Andrianavalomahefa}, BREAD \cite{2022Liu}, DOSUE \cite{2023Kotaka}, QUALIPHIDE \cite{2023Ramanathan}, BRASS-p \cite{2023Bajjali}, and Dark SRF \cite{2023Romanenko} utilize the non-resonant methods to achieve large bandwidths.
The signal-to-noise ratio (SNR) of a haloscope system is typically limited by factors
  such as the quality factor and volume of the cavity, the noise temperature and the data acquisition efficiency.
Plentiful efforts has been made to improve the SNR of the detection systems.
For example, HAYSTAC employed JPAs to minimize the noise \cite{2017Brubaker}, while QUAX used a superconductive cavity with a high quality factor to increase the potential dark matter signal \cite{2019Alesini}.
Groups like ADMX and CAPP fabricated cavities with large volumes to enhance the potential signal \cite{2022Cervantes1,2021Kwon}.
In order to further improve the searching ability, it is desirable to simultaneously push these key parameters to their limits.
However, this is practically challenging
  due to factors such as frequency drifts, difficulties in fabrication of the high-Q, large-volume cavity, dead time of electronic devices, and compatibility issues.

Here we focus on a promising region of the dark photon mass, which is in the order of tens of $\mathsf{\mu}$eV \cite{2012Arias,2016Graham,2016Ade}.
In order to improve the SNR, a superconducting cavity with a high quality factor and a large volume was utilized to maximize the potential dark photon signal.
To minimize the noise, a Josephson Parametric Amplifier (JPA) with a near-quantum-limited noise temperature was installed.
Additionally, we employed a homemade DAQ card based on the Field Programmable Gate Array (FPGA) to achieve a 100$\%$ duty cycle \cite{2020Tong}.
The hitherto highest sensitivity to $\chi$ is achieved in a 0.4 neV region around 26.96514 $\mu$eV.
While the constraints on $\chi$ have never touched the $10^{-16}$ line in prior searching efforts, this work sets an upper limit of $\chi$ of $6.0 \times 10^{-16}$ at 90$\%$ confidence level at the center of the region.


\begin{figure*}
  \centering
  \includegraphics[width=2\columnwidth]{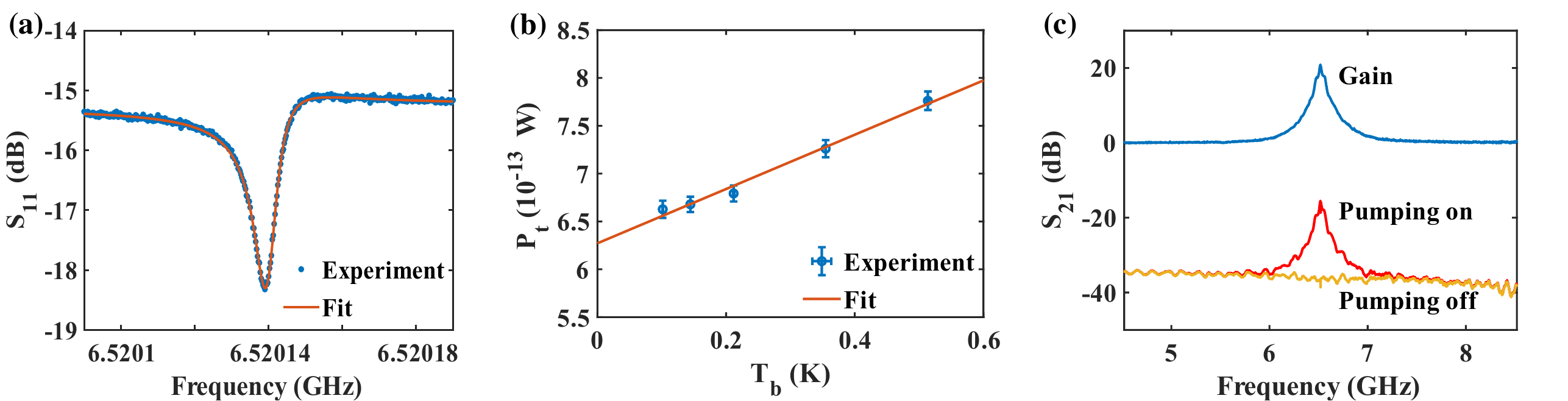}
  \caption{
      Calibration results.
      (a) Reflection coefficients of Port 1 of the cavity. The resonant frequency of the cavity, $f_{c}$, is 6.52014 GHz.
      (b) The relation between the noise power and the temperature of the 50 Ohm load. The noise temperature of the HEMT chain $T'_{\rm n}$ is 1.9 K.
      (c) Calibration result of the gain of the JPA.
          The red and yellow lines refer to the transmission coefficients S$_{21}$ from the Bypass line to the Output line when the JPA is on and off, respectively.
          Their difference is plotted as the blue line.
          The gain of the JPA at $f_{c}$ is $G_{\rm J} = $20 dB.
  }
  \label{Figure2}
\end{figure*}

Figure \ref{Figure1} shows the simplified diagram of the experimental setup (see more details in Sec. \uppercase\expandafter{\romannumeral1} in \cite{spp}),
  which consisted of two main parts:
  the detection part for searching dark photons and the calibration part for calibrating significant parameters of the detection part.
A temperature as low as 21 mK and extremely stable was achieved with a dilution refrigerator.

The detection part is presented in the green box in Fig. \ref{Figure1}.
It includes a superconducting cavity, an amplification chain, and a DAQ card.
The cylindrical superconducting cavity working at the TM$_{010}$ mode performed as an antenna to collect the dark photon signal.
Dark photons inside the cavity would continuously oscillate into ordinary photons due to the kinetic mixing and then generate a microwave (mw) signal.
The inset of Fig. \ref{Figure1} presents the picture of the superconducting cavity.
The cavity was made of 6061 aluminum alloy, whose critical temperature is around 1 K.
The inner side of the cavity was finely polished to enhance the Q factor.
The diameter and length are 35.31 mm and 150.00 mm, respectively.
The electric field of the TM$_{010}$ mode is along the cylindrical axis of the cavity, thus the filling factor can be expressed as
\begin{equation}
  C  = \frac{
              |\int dV \mathbf{E}|^{2}
            }
            {
              V\int dV |\mathbf{E}|^{2}
            }
            \langle\cos^{2}\theta\rangle ,
\label{C2}
\end{equation}
where $\theta$ stands for the angle between the polarization direction of the dark photon field and the electric field.
In the random polarization scenario, $\langle\cos\theta\rangle$ equals to 1/3 \cite{2021Caputo,2021Ghosh}.
The simulation result shows that the filling factor $C$ is 0.23.
There were two ports in the cavity: the strong one (Port 1) for readout and the weak one (Port 2) for injection of a simulated dark photon signal.
The coupling from the cavity to the transmission lines was achieved through coaxial probes.
In order to avoid direct coupling between the two probes, they were fabricated on the opposite sides of the cavity.
The amplification chain was connected to the Port 1.
A JPA accompanied by a circulator was taken as the first-stage amplifier since its noise temperature approaches the quantum limit $T_{n} = \hbar\omega/k_{B}$ \cite{1982Caves}.
The signal was further amplified by High-Electron-Mobility Transistor (HEMT) amplifiers to a level where the electrical noise doesn't dominate,
  and readout by a homemade DAQ card based on FPGAs (see more details in Sec. \uppercase\expandafter{\romannumeral2} in \cite{spp}).
The parallel sampling module and Fast-Fourier-Transform (FFT) modules of the DAQ card allowed for
  simultaneous data collection and frequency spectrum calculation with a 100$\%$ duty cycle.

The calibration part is indicated by the blue box in Fig. \ref{Figure1}.
It aimed to measure significant parameters of the detection part, including the resonant frequency of the cavity and the noise temperature of the HEMTs.
The Strong line allowed for the measurement of the resonant frequency of the cavity.
The 50 Ohm load and a heater (not shown) played the role of a temperature-variable black radiation source,
  and a low-temperature switch supported convenient shift between the cavity and the 50 Ohm load.
They were used to measure the noise temperature of the HEMTs through the Y-factor method.
The Bypass line was used to measure the temperature noise and the gain of the JPA.
The Weak line was used for injection of a simulated dark photon signal verification of the detection system (see more details in Sec. \uppercase\expandafter{\romannumeral3} in \cite{spp}).


To locate the resonant frequency $f_{c}$ of the cavity, the reflection coefficient of Port 1 of the cavity was measured, as presented in Fig \ref{Figure2}(a).
The result is fitted to the equation \cite{2020Kudra}
\begin{equation}
  S_{11}  = \alpha e^{i\psi} [1-\frac{2\beta_{1}/(1+\beta_{1})e^{i\phi}}{1+2iQ_{L}(f/f_{c}-1)}],
\label{S11}
\end{equation}
  where $\alpha$ and $\psi$ are the power loss and the phase shift during transmission, respectively,
  and the parameter $\phi$ describes the impedance mismatch, which makes the spectrum asymmetric.
The fitting gives $f_{c} = $ 6.52014 GHz.

During the calibration of the HEMTs, the pumping of the JPA was off, and the switch was thrown to the 50 Ohm load.
The thermal noise power of the system can be expressed as
\begin{equation}
  P_{\rm t}  = G_{\rm H}B[hf_{c}(\frac{1}{e^{hf_{c}/k_{\rm B}T'_{\rm n}}-1}+\frac{1}{2})+k_{\rm B}T_{\rm b}],
\label{Y-factor}
\end{equation}
where $G_{\rm H}$ is the gain of the HEMT chain, $k_{\rm B}$ is the Boltzmann constant,
  $B$ is the bin width of the spectrum, $T'_{\rm n}$ is the noise temperature
  of the HEMT chain, and $T_{\rm b}$ is the physical temperature of the 50 Ohm load.
By varying $T_{\rm b}$ and fitting the temperature dependence of $P_{\rm t}$, as plotted in Fig \ref{Figure2}(b),
  $G_{\rm H}=76.9 \pm 0.3$ dB and $T'_{\rm n} = 1.9\pm0.1$ K are obtained.

The transmission coefficients from the Bypass line to the Output line
  as the pumping of the JPA was on and off are shown as the red and yellow lines in Fig. \ref{Figure2}(c), respectively.
The blue line refers to their difference, i.e., the gain of the JPA, which is 20 dB at the resonant frequency of the cavity.
The noise temperature of the JPA $T_{\rm n}$ is calculated to be $297 \pm 22$ mK from the formula:
\begin{equation}
  T_{\rm n}  = \frac{1}{G_{\rm J}}\frac{P_{\rm J\_on}}{P_{\rm J\_off}}T'_{\rm n},
\label{T_noise}
\end{equation}
  where $G_{\rm J}$ is the gain of the JPA,
  $P_{\rm J\_on}$ and $P_{\rm J\_off}$ are the thermal noise power measured from the Output line with the JPA on and off, respectively.
Since the bandwidth of the JPA is much broader than that of the cavity, its gain and noise temperature can be considered uniform in the detection region.

\begin{figure}
\centering
\includegraphics[width=1\columnwidth]{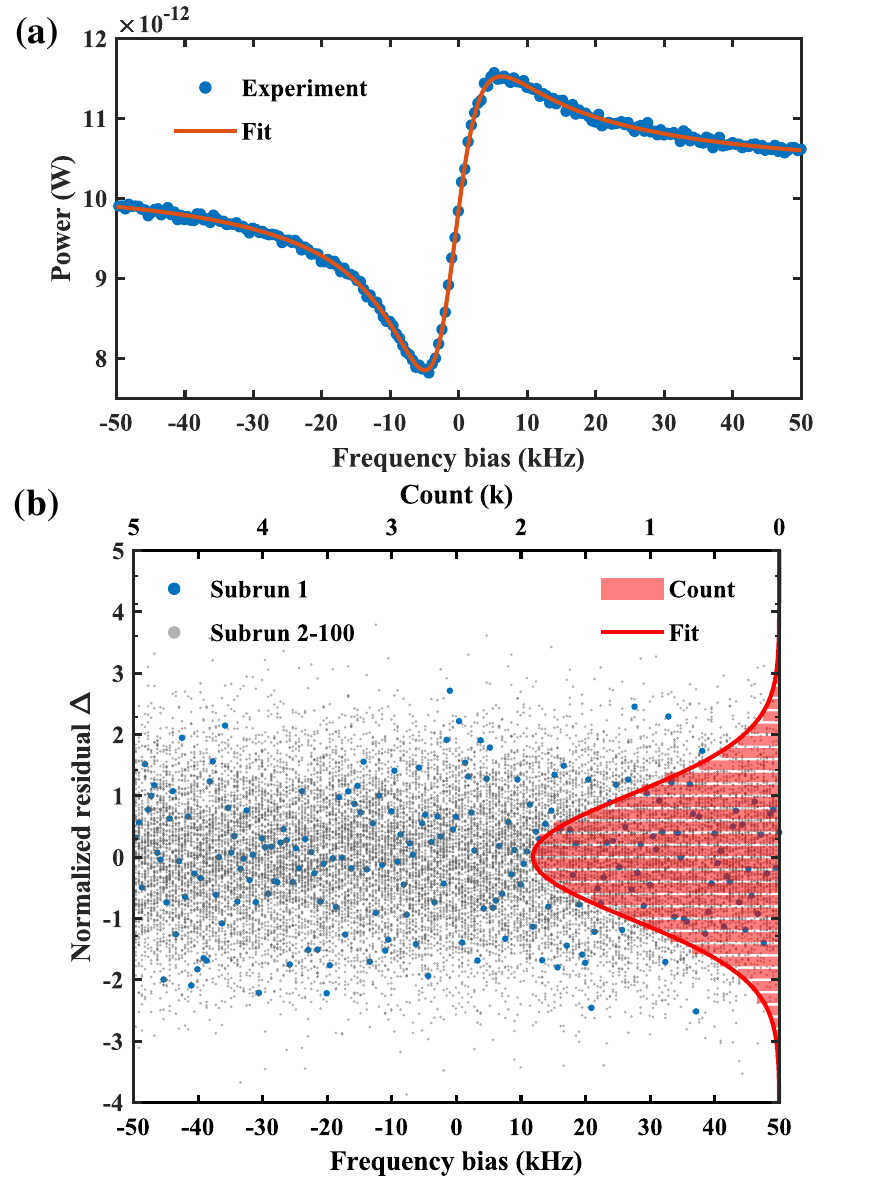}
\caption{
  (a) The power spectrum of the first subrun.
      The blue dots are the measured powers averaged from 400 raw spectra.
      Two consecutive fittings were used to remove the baseline due to the cavity and the mw devices.
      The product of the fitting results is shown as the orange line.
  (b) The normalized residuals $\Delta$ and their distribution.
      The normalized residuals obtained from subrun 1 and subrun 2-99 are plotted as the blue dots
        and the gray dots, respectively.
      The red bars are the counts of data while the red line is the fitting result which is pretty close to the
        unit Gaussian distribution.
}
\label{Figure3}
\end{figure}

During the detection of dark photons, all the inputs except the pumping of the JPA were off.
The data collection lasted for six hours.
Forty thousand raw spectra have been obtained.
The frequency bin width $B$ of each spectrum was 0.48 kHz.
These raw spectra were divided into 100 subruns, each containing 400 raw spectra.
The average of the first subrun is shown as the blue dots in Fig \ref{Figure3} (a).
The line shape is right the resonant dip of the cavity.

The spectra are firstly fitted to equation \ref{S11} (Fit$_{1}$) to remove the baseline due to the cavity.
Then, the fluctuations in the transmission efficiency of the devices are removed by applying a sum-of-sines fitting (Fit$_{2}$).
Meanwhile, the values of the coupling coefficient $\beta_{1}$ and the quality factor of the cavity $Q_{L}$ during each subrun are obtained from the fitting results.
The product of the two fitting results of the first subrun is shown as the orange line in Fig \ref{Figure3} (a), with $\beta_{1} = 0.4$ and $Q_{L} = 6.1\times10^{5}$.
The blue dots in Fig \ref{Figure3} (b) depict the standardized power residuals $\Delta$ of subrun 1,
\begin{equation}
  \Delta = \frac{\Delta_{\rm s}}{\sigma_{\rm s}},
  \label{Delta}
\end{equation}
where $\Delta_{\rm s}$ is the normalized power excess obtained from the baseline removing operation, and $\sigma_{\rm s}$ is the standard deviation of the normalized power excess (see more details in Sec. \uppercase\expandafter{\romannumeral4} in \cite{spp}).
The gray dots depict the standardized residuals of subrun 2-100.

A power excess over five times the standard deviation is required for a data point to be claimed as a candidate signal.
To check the presence of any candidate signal in our results, the standardized power residuals are counted,
  as illustrated by the red bars in Fig. \ref{Figure3} (b).
The red line is the fitting result with the Gaussian distribution N$(\mu,\sigma'^{2})$,
  where $\mu$ is the mean value of the data and $\sigma'$ is the standard deviation.
The distribution of the standardized power residuals fits well with the unit Gaussian distribution, with $\mu = 0.009 \pm 0.010$ and $\sigma' = 1.009 \pm 0.010$.
No power excess over 5$\sigma$ is observed.
Therefore, our results provide an exclusion of dark photons in the detection bandwidth.

\begin{figure}
  \centering
  \includegraphics[width=1\columnwidth]{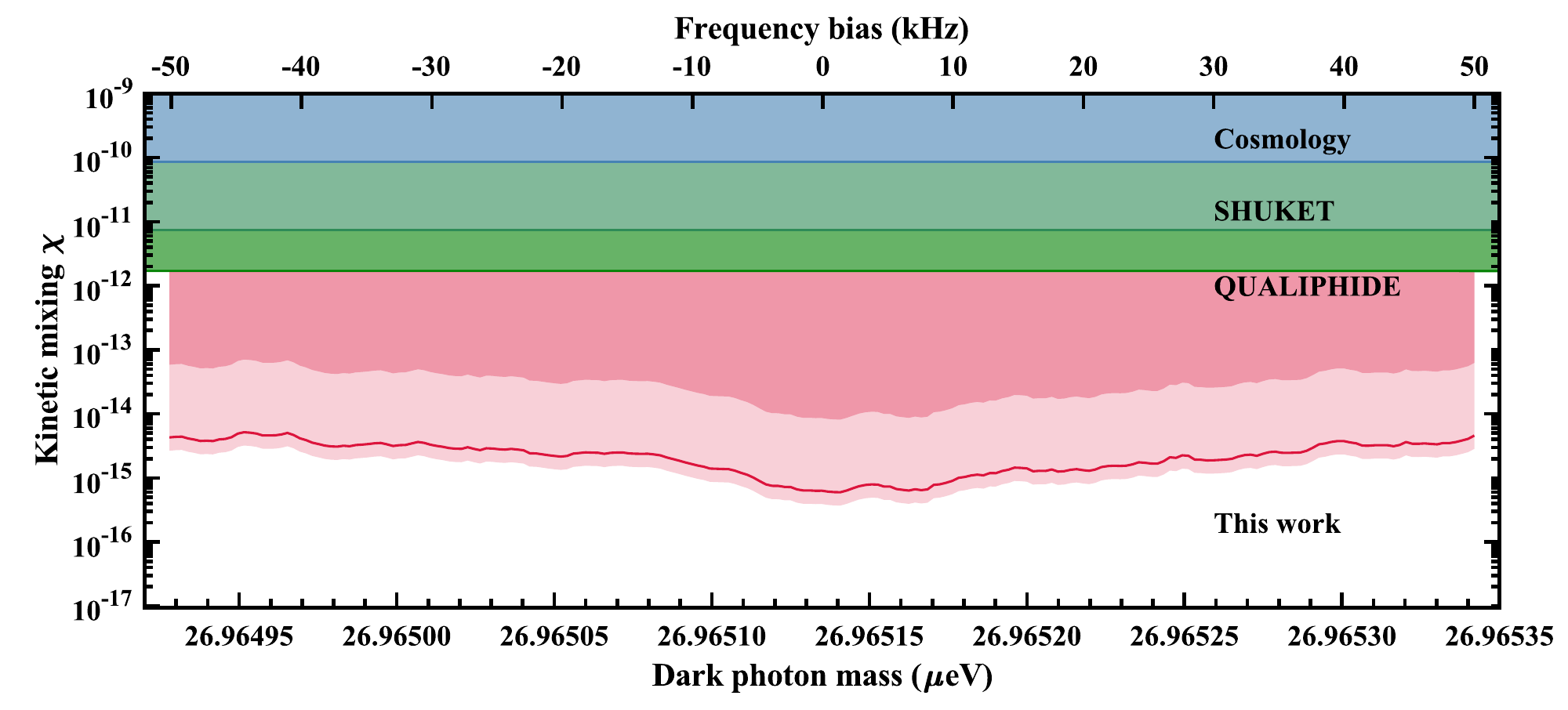}
  \caption{
    Upper limits on the kinetic mixing between dark photons and ordinary photons.
    The red line refers to the upper bounds of the kinetic mixing in the random polarization scenario set by this work.
        The shallow and dark pink region refer to the excluded parameter space in the linear polarization scenario in the best and worst condition, respectively.
        The x-ticks at the top are the frequency biases from the central value, 6.52014 GHz,
          while the x-ticks at the bottom are the corresponding dark photon masses.
        The blue, shallow green, and deep green regions refer to the previous results of
          cosmological observations \cite{2012Arias}, the SHUKET \cite{2019Brun} and the QUALIPHIDE \cite{2023Ramanathan} experiments, respectively.
        Data adapted from \cite{2020Hare}.
    }
  \label{Figure4}
\end{figure}

The data analysis that provides the constraints of the kinetic mixing $\chi$ follows the method developed in the ADMX experiment \cite{2022Cervantes2} (see more details in Sec. \uppercase\expandafter{\romannumeral4}, \uppercase\expandafter{\romannumeral5} in \cite{spp}).
Figure \ref{Figure4} shows the upper bounds on the kinetic mixing
  in the range of dark photon masses from 26.96493 $\mu$eV to 26.96534 $\mu$eV
  with a confidence level of 90$\%$.
The red line refers to the upper bounds of the kinetic mixing in the random polarization scenario.
An upper bound of $\chi < 6.0\times10^{-16}$ is achieved at 26.96514 $\mu$eV, which is
  more than three orders of magnitude more stringent than the bound established by the QUALIPHIDE experiment.
For a linear polarization scenario, the constraints depend on the angle between the polarization direction of the dark photon field and the electric field \cite{2021Caputo}.
Although the polarization of dark photons is unknown, the best constraints and the worst constraints can be obtained by going over all the possible polarization directions,
  as the shallow pink and the dark pink region in Fig. \ref{Figure4} shows.

In summary, we report an experimental search for dark photons, which is a possible candidate for dark matter.
Our results improve the constraints of the kinetic mixing by more than three orders of magnitude.
This work has shown an outstanding $\frac{Q_{L}Q_{a}}{Q_{L}+Q_{a}}VC\frac{\beta}{1+\beta}$ that is comparable with existing resonant haloscopes.
For example, we have achieved an improvement of around 2 times comparing with QUAX \cite{2022Alesini} and HAYSTAC \cite{2017Brubaker}.
However, this work is at a preliminary stage because of the lacking of a tuning system.
There is a large room for improvement of $\frac{Q_{L}Q_{a}}{Q_{L}+Q_{a}}V^{2}C^{2}\frac{\beta^{2}}{(1+\beta)^{2}}$,
  which is more concerned in terms of scan rate.
For future dark matter search experiments, new tunable cavities with larger mode volume are under construction.
The sensitivity of our apparatus can be further improved by taking advantage of the single mw photon counting technique \cite{2016Inomata,2018Royer,2023Balembois}.
The searching efficiency can be multiplied by combining a series of cavities with different resonant frequencies.
The experimental setup can also be used to test other predictions arising from new physics beyond the Standard Model, such as the existence of axions and axion-like particles, etc., with further upgrades \cite{2020Luzio,2021Sikivie}.
In the future, it is practicable to examine the Kim–Shifman–Vainshtein–Zakharov model with the present setup, a superconducting cavity with through holes and 6 T magnet \cite{1979Kim,1980Shifman}.
The sensitivity required by the Dine-Fischler-Srednicki-Zhitnitsky model is also within reach \cite{1980Zhitnitsky,1981Dine}.

The authors are grateful to the anonymous referees for their valuable comments.
This work was supported by the Innovation Program for Quantum Science and Technology (2021ZD0302200), the Chinese Academy of Sciences (No. GJJSTD20200001), the National Key R\&D Program of China (Grant No.2021YFB3202800, No. 2018YFA0306600, No. 2021YFC2203100), Anhui Initiative in Quantum Information Technologies (Grant No. AHY050000), NSFC (12150010, 12205290, 12261160569, 12261131497).  Y. F. C. and M. J.  thank the Fundamental Research Funds for Central Universities. Y.F.C. is supported in part by CAS Young Interdisciplinary Innovation Team (JCTD-2022-20), 111 Project (B23042). M. J. Thanks is supported in part by China Postdoctoral Science Foundation (2022TQ0330). This work was partially carried out at the USTC Center for Micro and Nanoscale Research and Fabrication.

\appendix

\section{The Schematic Diagram of the Experimental Apparatus}

\begin{figure*}[htp]
  \centering
  \includegraphics[width=2\columnwidth]{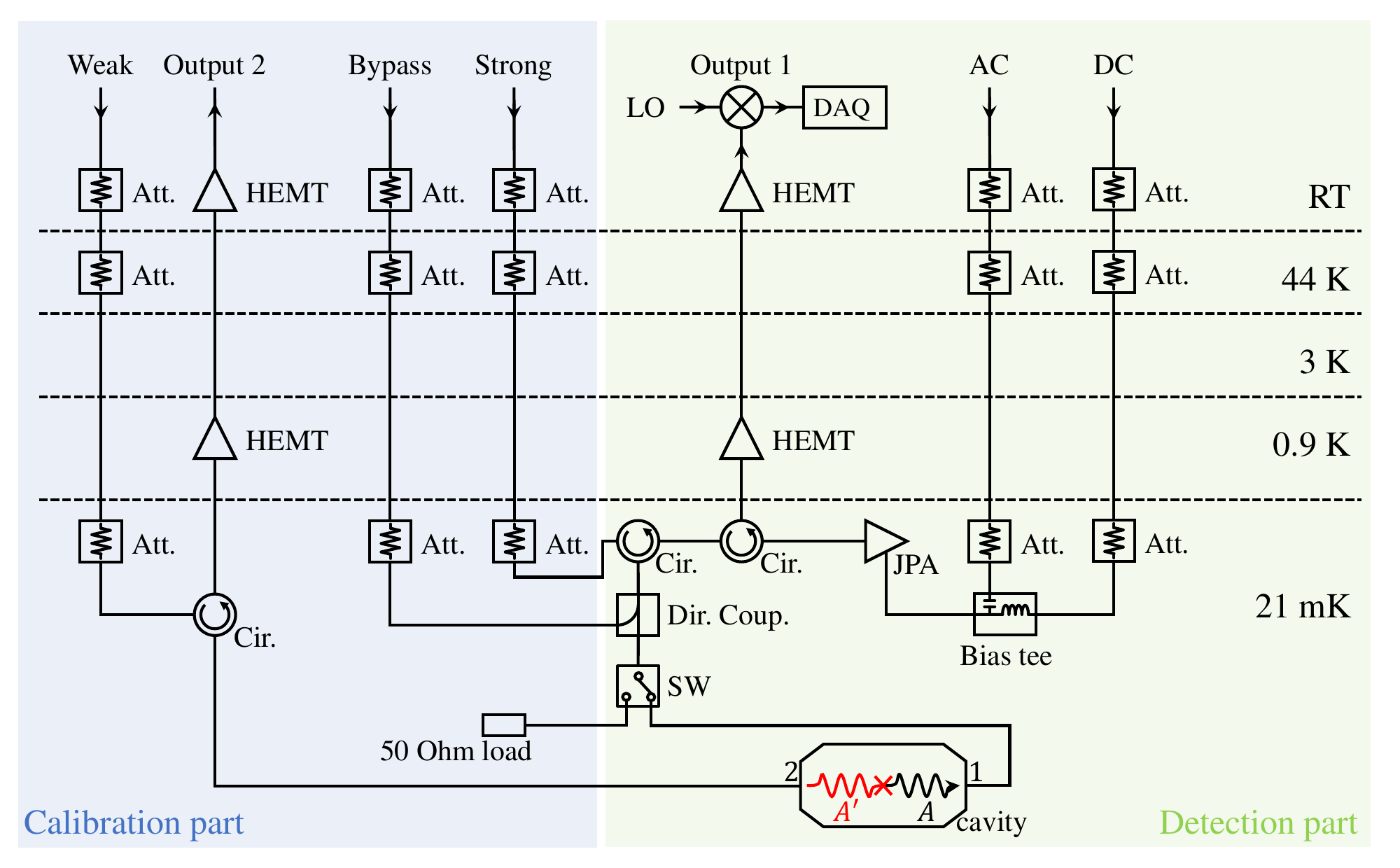}
  \caption{
    Schematic diagram of the experimental apparatus.}
  \label{FigureS1}
\end{figure*}

The schematic diagram of the experimental apparatus is shown in Fig. \ref{FigureS1}.
The whole setup was housed in a multistage cryogenic system.
Each horizontal dashed line in Fig. \ref{FigureS1} refers to a cold flange.
A temperature as low as 21 mK was achieved in the last stage with a dilution refrigerator.

In the detection part, two input lines (AC and DC) were used for the pumping of the JPA.
A bias current of a few milliamperes was injected from the DC line, while a mw signal with twice the frequency as the working point was injected from the AC line.
The signal from the cavity was first reflected and amplified by the JPA, and then further amplified by a cascade of HEMTs.
Then, the amplified signal was mixed with a local oscillator signal and down-converted to 1 MHz.
The down-converted signal was finally acquired by a homemade DAQ card based on FPGAs.

Three input lines were employed in the calibration part.
By measuring the transmission coefficients from the Strong and the Bypass line to the Output line 1,
  $S_{11}$ of the superconducting cavity and the gain curve of the JPA could be obtained, respectively.
The Weak line was used for the injection of a simulated dark photon signal.
It was accompanied by another Output line for the measurement of the reflection coefficients of Port 2.

Each input line was carefully attenuated so that the noise was suppressed to a negligible extent.

\section{The Data Acquisition Card}

\begin{figure}[htp]
  \centering
  \includegraphics[width=1\columnwidth]{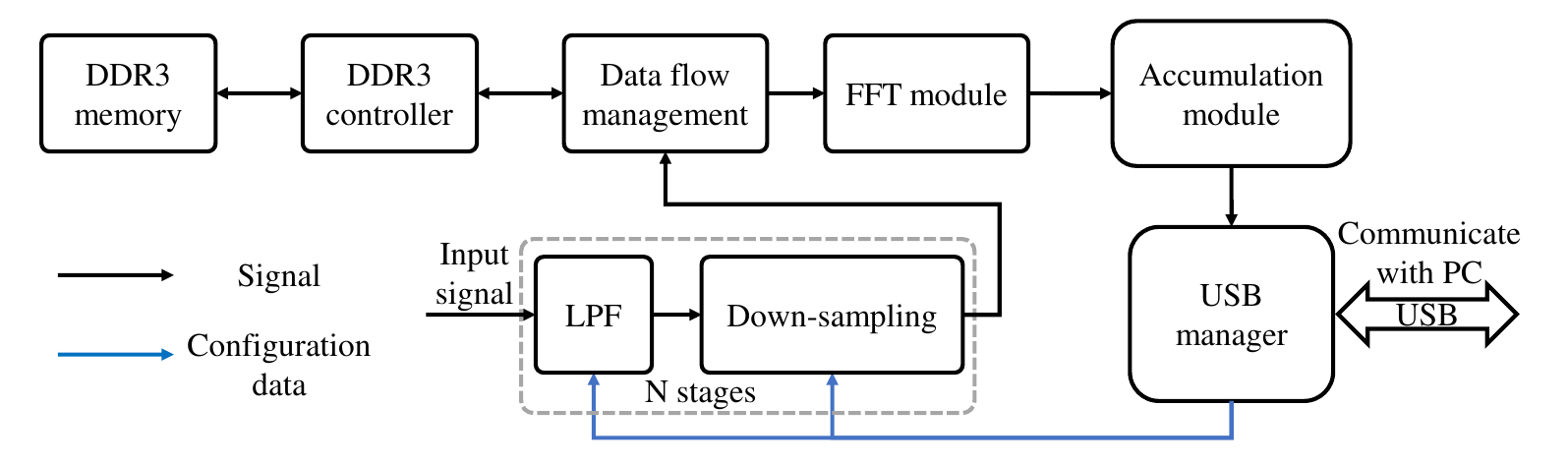}
  \caption{
    Simplified schematic diagram of the DAQ card.}
  \label{FigureS2}
\end{figure}

A data acquisition (DAQ) card based on the Field Programmable Gate Array (FPGA) has been developed to achieve a 100$\%$ duty cycle.
Figure. \ref{FigureS2} presents the simplified schematic diagram of the DAQ card.
The down-converted input digital signals are processed by an N-stage digital Low-Pass Filter (LPF) to suppress the noise and the aliasing, and an N-stage downsampling is used to improve the frequency resolution, where N is a configurable parameter.
In this work, N was set to 6.
In order to realize a 100$\%$ real-time data utilization ratio for the data flow, a Double-DataRate-III (DDR3) memory is used to provide sufficient on-board data storage capacity.
The time-domain data are temporarily stored in the DDR3 before being sent to the FFT module.
The high parallel processing capability of the FPGA allows the collection of data and the calculation of frequency spectra to be conducted simultaneously.
The output spectra are averaged in an accumulation module and the average result is sent to the host computer through a high-speed universal serial bus (USB) port.
More details about the DAQ card can be seen in \cite{2020Tong}.

\begin{figure}[htp]
  \centering
  \includegraphics[width=1\columnwidth]{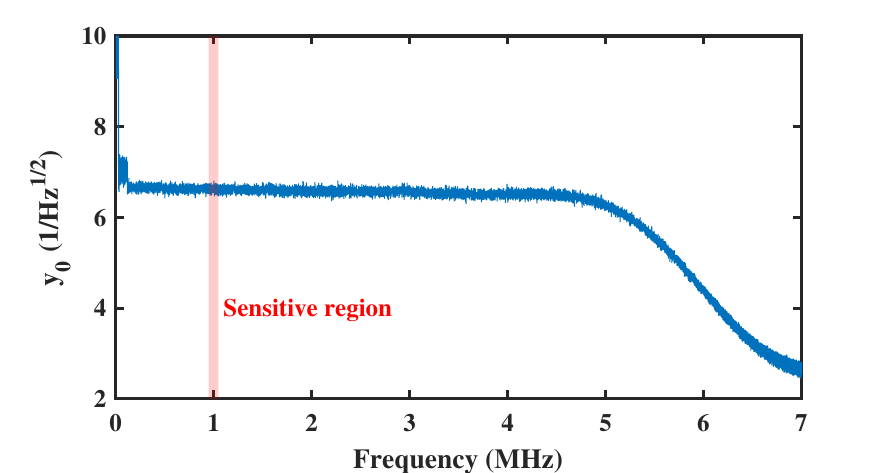}
  \caption{
    Electronic noise power spectrum of the DAQ card.
        The sensitive region is indicated by the red ribbon.}
  \label{FigureS3}
\end{figure}

To test the electrical noise of the DAQ card, the output spectrum without any input signal was measured, as shown in Fig. \ref{FigureS3}.
The outputs $y$ of the DAQ card are values in the unit of $1/$Hz$^{1/2}$ that are proportional to the actual input voltages.
A noise floor in the frequency range from 0 MHz to 5 MHz is observed.
The sharp decrease of the electrical noise beyond 5 MHz is caused by the multi-stage LPF.
Actually, only a narrow region around 1 MHz is concerned since it corresponds to the frequency band that is sensitive to the dark photon.
The electrical noise floor in this region is 6.6 Hz$^{-1/2}$.

\begin{figure}[htp]
  \centering
  \includegraphics[width=1\columnwidth]{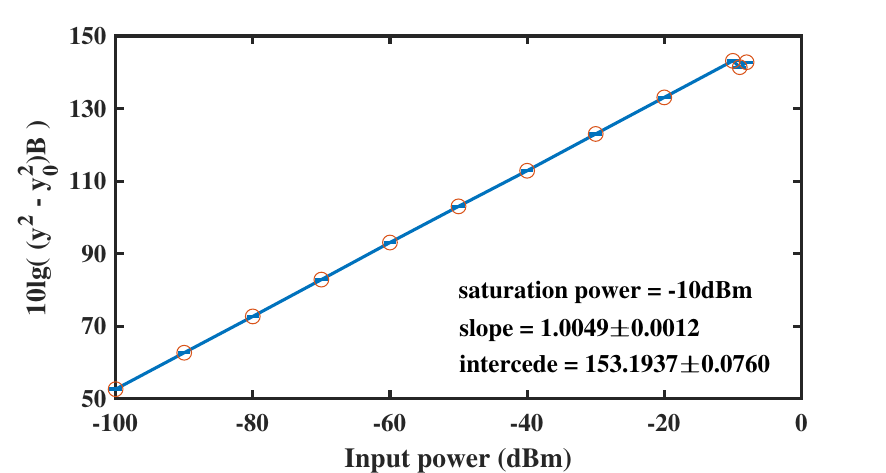}
  \caption{
    Relationship between the output values of the DAQ card and the input powers.}
  \label{FigureS4}
\end{figure}

In order to map the outputs of the DAQ card to the actual power, the card is calibrated by inputting mw signals with known power levels.
Fig. \ref{FigureS4} shows the relationship between the output values and the input signals with power levels ranging from -100 to -8 dBm.
The result is fitted to the equation
\begin{equation}
  10\lg[(y^{2}-y_0{^{2}})B]  = 10\lg\xi + \widetilde{P} _{\rm in},
\label{DAQ1}
\end{equation}
  which is equivalent to
\begin{equation}
  (y^{2}-y_0{^{2}})B  = \xi P_{\rm in},
  \label{DAQ2}
\end{equation}
where $y_{0}$ is the electrical noise, $B$ is the bin width of the spectrum, $\xi$ is the conversion coefficient, $P_{\rm in}$ and $\widetilde{P} _{\rm in}$ are the power of the input signal in the unit of mW and dBm, respectively.
The fitting result gives $\xi = 10^{15.32}$ mW$^{-1}$.
Note that the relation is no longer linear when the input power exceeds -10 dBm, since the power is so high that the DAQ card will be saturated.

\section{Verification of the detection system}

\begin{figure}[htp]
  \centering
  \includegraphics[width=1\columnwidth]{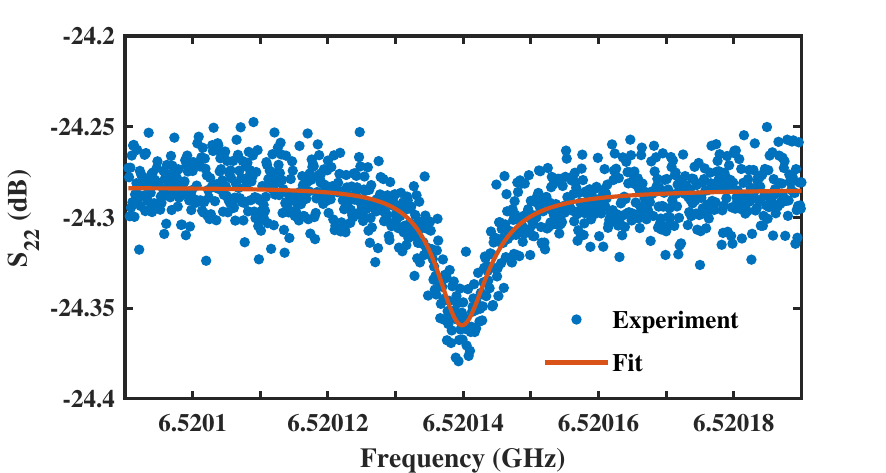}
  \caption{The reflection coefficient of the Port 2 of the cavity. The coupling coefficient $\beta_{2}$ is 0.0043.}
  \label{FigureS5}
\end{figure}

To demonstrate the capability of our system in detecting a potential weak dark photon signal, the cavity was designed to have an extra port (Port 2) for injection of a simulated dark photon signal.
The coupling coefficient of the Port 2 was adjusted to a very small value.
Figure \ref{FigureS5} shows the reflection coefficient of the Port 2.
The fitting gives $\beta_{2} = 0.0043$.

\begin{figure}[htp]
  \centering
  \includegraphics[width=1\columnwidth]{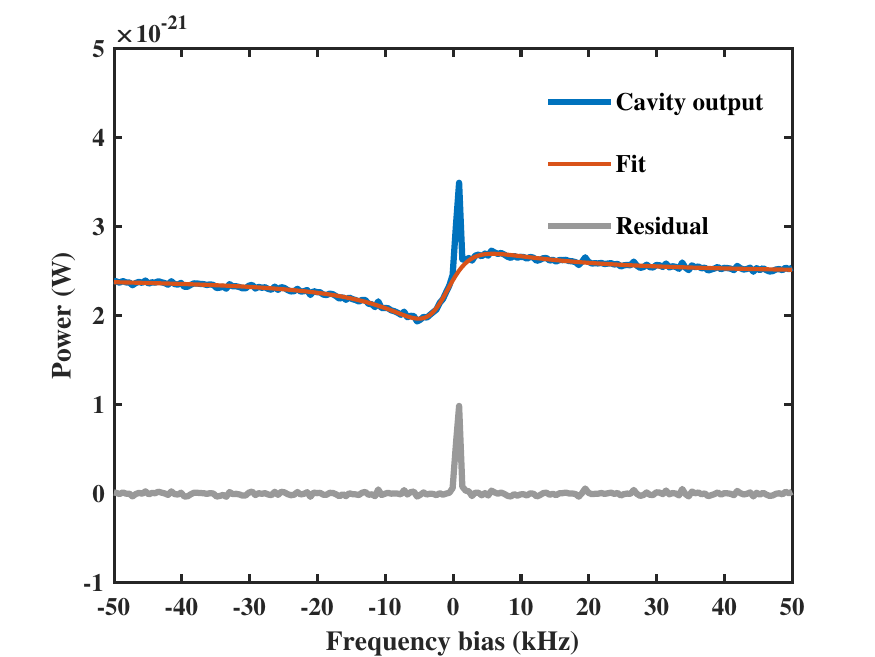}
  \caption{
    The result of the simulated dark photon experiment.
    The blue line is the output power spectrum at the Port 1 of the cavity.
    The orange line is the baseline of the spectrum.
    Their difference is plotted as the gray line.}
  \label{FigureS6}
\end{figure}

A pure-tone mw signal with the power being -43 dBm was synthesized to simulate the dark photon signal.
The power was further attenuated to -181 dBm, i.e., one zeptowatt, corresponding to a dark photon signal with $\chi = 6\times 10^{-15}$.
The blue and orange lines in Fig. \ref{FigureS6} represent the power spectrum averaged over one minute at the Port 1 of the cavity and the baseline of the spectrum, respectively.
By subtracting the baseline from the power spectrum, the power residual can be obtained, as shown by the gray line.
We do observe a simulated dark photon signal, whose amplitude is consistent with the input signal.
The signal-to-noise ratio (SNR) is 58.7.

\section{Data analysis}

\begin{figure}[htp]
  \centering
  \includegraphics[width=1\columnwidth]{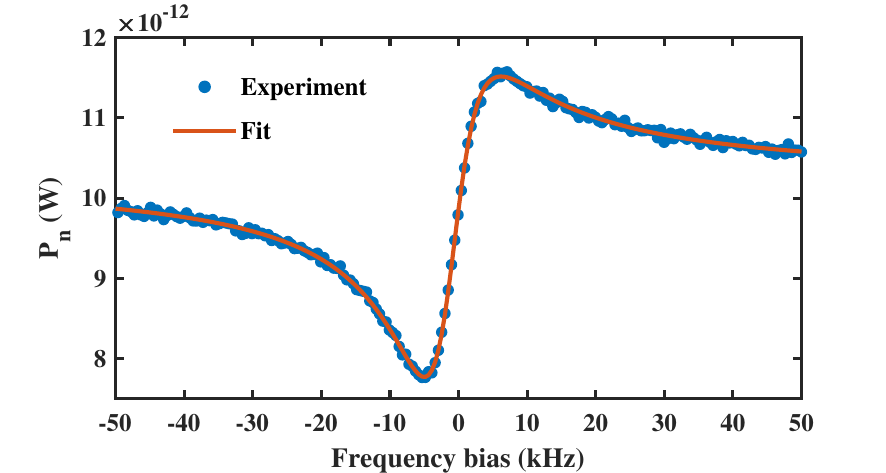}
  \caption{
    The average power spectrum of subrun 1.
    The blue dots stand for the mean value of 400 raw spectra, and the orange line stands for the fitting to the $S_{11}$ function of the cavity.
    }
  \label{FigureS7}
\end{figure}

\begin{figure*}[htp]
  \centering
  \includegraphics[width=2\columnwidth]{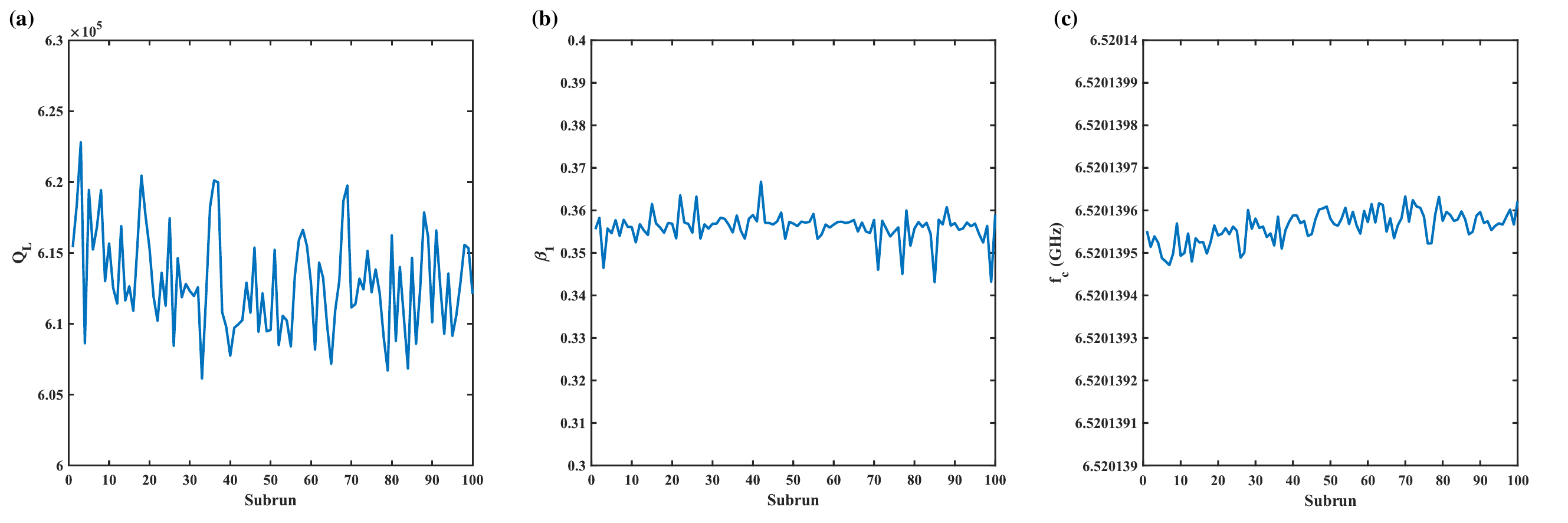}
  \caption{
    Fluctuations of experimental parameters during the data collection.
    (a) The quality factor.
    (b) The coupling coefficient.
    (c) The resonant frequency.
    }
  \label{FigureS8}
\end{figure*}

After the six-hour data collection, forty thousand raw spectra have been obtained.
In order to suppress the noise, the raw spectra are divided into 100 subruns.
Each subrun contains 400 neighbouring raw spectra, and the average of them is calculated.
The average result of the first subrun is shown as the blue dots in Fig. \ref{FigureS7}.

The curve is due to the cavity.
Since there was a static temperature gradient between the cavity and the environment, the power spectrum can be expressed as
\begin{equation}
    P_{\rm n}  \varpropto T_{\rm n} + T_{\rm e} + (T_{\rm c}-T_{\rm e})|S_{11}|^{2},
  \label{T_noise}
\end{equation}
  where $T_{\rm e}$ is the temperature of the environment, $T_{\rm c}$ is the temperature of the cavity, and $S_{11}$ is the reflection coefficient of the cavity (Equation 4 in the main text).
The fitting result Fit$_{1}$ is shown as the orange line in Fig. \ref{FigureS7}.
The quality factor $Q_{\rm L}$, the coupling coefficient $\beta_{1}$, and the resonant frequency $f_{\rm c}$ for each subrun are also accessible through the fitting results.
Their fluctuations during the data collection are shown in Fig. \ref{FigureS8}.

\begin{figure}[htp]
  \centering
  \includegraphics[width=1\columnwidth]{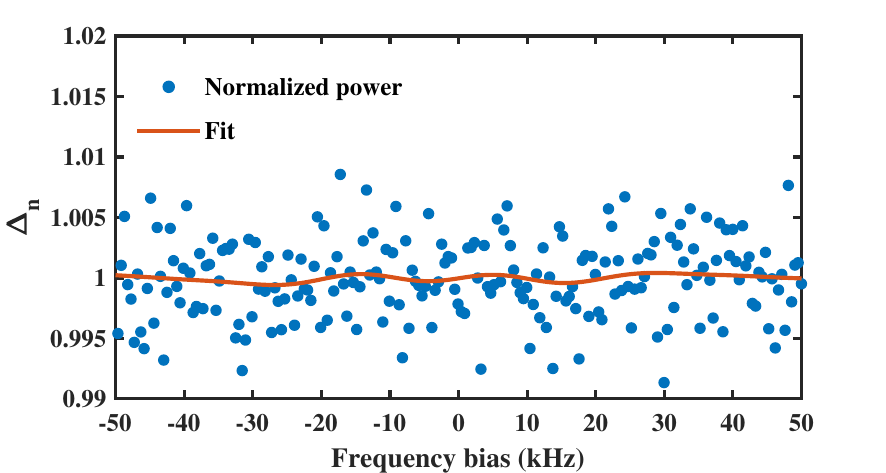}
  \caption{
    The normalized power spectrum of subrun 1.
    The blue dots stand for the power spectra divided by the cavity baseline
    and the orange line stands for the fitting to a sum of sinusoidal functions.
    }
  \label{FigureS9}
\end{figure}

\begin{table}
  \linespread{1.5}
   \caption{Fitting parameters in Fig. \ref{FigureS9}}
   \label{table1}
   \begin{tabular}{p{1cm} p{2cm} p{2cm} p{0.5cm}}
    \hline
    \hline
      $k$  & $C_{k}$ & $\nu_{k}$ & $\xi_{k}$\\
    \hline
      {0}  & $1.0$                  & -    & - \\
      {1}  & $0.2\times10^{-3}$     & 3.6  & 3.0 \\
      {2}  & $1.2\times10^{-3}$     & 10.0 & 2.9 \\
      {3}  & $0.1\times10^{-3}$     & 2.8  & 3.0 \\
      {4}  & $-1.3\times10^{-3}$    & 9.1  & 2.9 \\
      \hline
      \hline
    \end{tabular}
  \label{Table1}
\end{table}

\begin{figure}[htp]
  \centering
  \includegraphics[width=1\columnwidth]{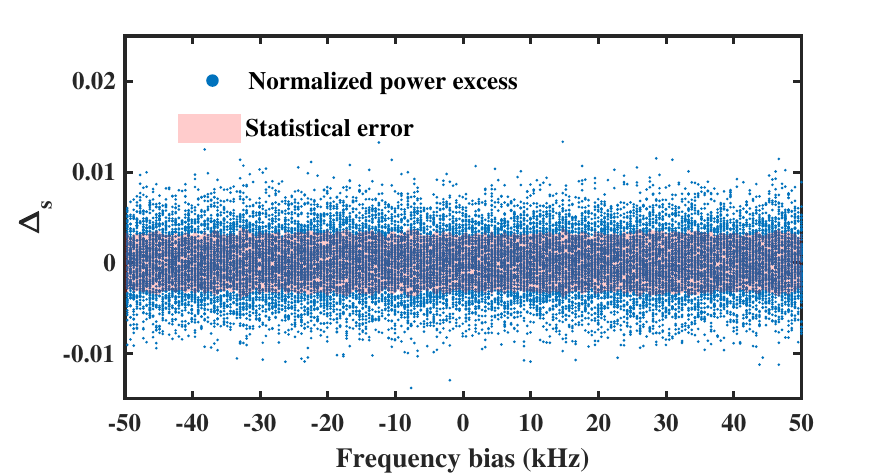}
  \caption{
    The normalized power excess of all subruns.
    The shallow red ribbon stands for the statistical error.
    }
  \label{FigureS10}
\end{figure}

Every averaged power spectrum is divided by the cavity baseline $\rm Fit_{1}$,
\begin{equation}
    \Delta_{\rm n}  = \frac{P_{\rm n}}{\rm Fit_{1}}.
  \label{T_noise}
\end{equation}
Figure \ref{FigureS9} shows the normalized power spectrum $\Delta_{\rm n}$ of the first subrun.
The fluctuations in the normalized power spectrum are caused by the variation in transmission efficiency of the mw devices.
They are removed by fitting the normalized power spectrum to a sum of sines,
\begin{equation}
  {\rm Fit_{2}}  = C_{0} + \sum_{k = 1}^{4} C_{k} \sin(\frac{f}{\nu_{k}}+\xi_{k}),
  \label{T_noise}
\end{equation}
as the orange line in Fig. \ref{FigureS9} shows.
The fitting parameters in this step are shown in Table \ref{Table1}.
By dividing the normalized power spectrum by this fitting result and subtracting the central value of 1 from the quotient, we obtain the normalized power excess
\begin{equation}
    \Delta_{\rm s}  = \frac{\Delta_{\rm n}}{\rm Fit_{2}} - 1,
  \label{T_noise}
\end{equation}
  where $\rm Fit_{2}$ is the fitting result of the variation of transmission efficiency.
The same operations are also performed on the rescaled power spectra of subrun 2-100.
The blue dots in Fig. \ref{FigureS10} show the normalized power excess obtained from all subruns.
The standard deviation is taken as the statistical error, as the shallow red ribbon shows.

\begin{figure}[htp]
  \centering
  \includegraphics[width=1\columnwidth]{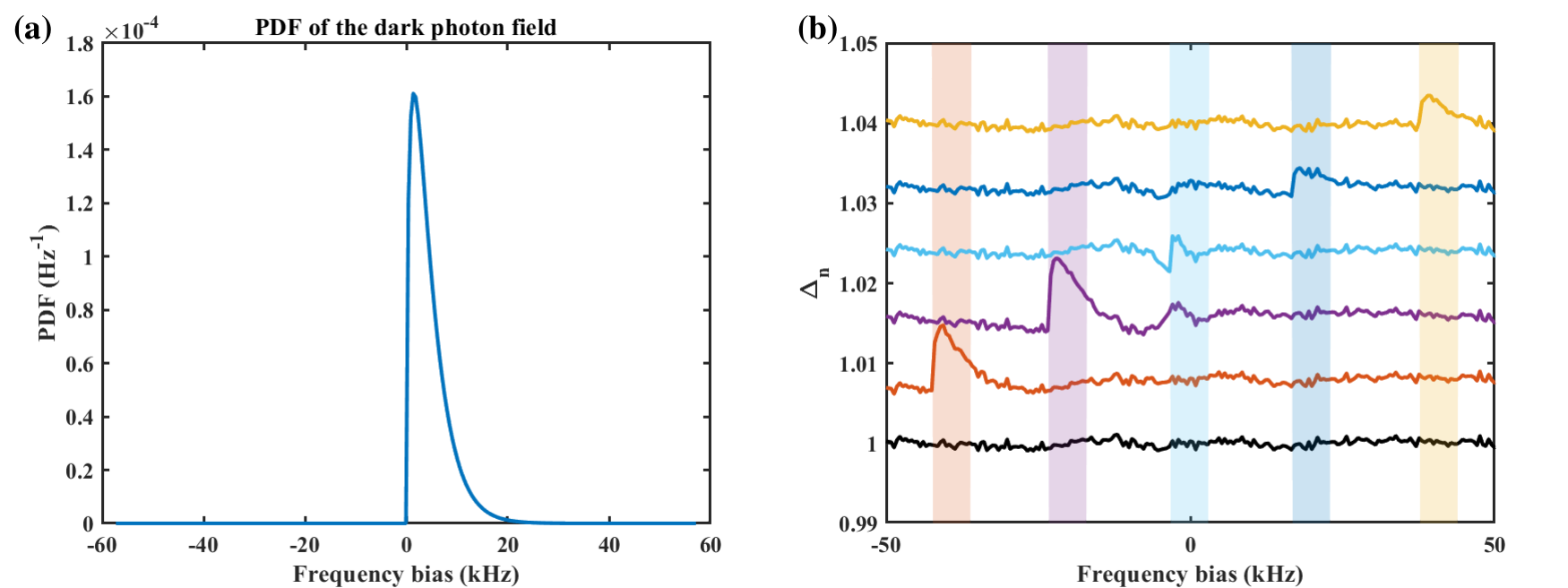}
  \caption{
    (a) The expected line shape of dark photons.
    (b) Normalized power spectra with simulated dark photon signals added to the original spectrum.
    }
  \label{FigureS11}
\end{figure}

The frequency distribution of the dark photon can be described by the Maxwell-Boltzmann distribution,
\begin{equation}
    \mathcal{F}(f) = 2\sqrt{\frac{f-f_{\rm a}}{\pi}}[\frac{3}{\eta f_{\rm a}\langle(v_{\rm a}/c)^{2}\rangle}]^{\frac{3}{2}}\exp\frac{-3(f-f_{\rm a})}{\eta f_{\rm a}\langle(v_{\rm a}/c)^{2}\rangle},
  \label{T_noise}
\end{equation}
  where $f_{\rm a}$ is the frequency of the photon that couples to the dark photon, $c$ is the velocity of light, $v_{\rm a}$ is the velocity of the dark photon, and $\eta$ is a correction factor introduced by the complex motion of the lab frame.
Here $\langle v_{\rm a}^{2}\rangle = 1.5\langle v_{\rm a}\rangle^{2} = (270 \rm km/s)^{2}$ and $\eta=1.7$ is adopted \cite{1985Krauss,1990Turner,2017Brubaker}.
For the dark photon in the frequency region around 6.52 GHz, the quality factor $Q_{a}$ and linewidth $\Delta f_{a}$ are $1\times 10^{6}$ and 6.52 kHz, respectively.
Figure \ref{FigureS11}(a) shows the expected line shape of the dark photon signal.
Given that the fitting results give $Q_{\rm L} < Q_{\rm a}$ and $\Delta f_{a}/{\nu_{k}} < \pi$,
  and that the line shape of dark photons (Maxwell-Boltzmann) is quite different from that of the cavity (Lorentzian),
  if a dark photon signal existed, it would survive in the baseline removing operations and show up in the normalized power excess spectrum as a narrow peak.

To further demonstrate this point, we performed the first baseline removal operation with simulated dark photon signals added to the original spectrum.
Figure \ref{FigureS11}(b) shows 5 examples of the results after the first baseline removal.
The strength of the signals is set to 5 times of our final constraints at the corresponding frequencies.
The black line refers to the original data (no dark photon signal),
  while the colored lines refer to the data with dark photon signals at different frequencies.
The ribbons of corresponding colors indicate the positions of the dark photon signals.
The signals are obvious, especially for the sharp rising edges.
We have also compared the amplitude of the signals after the baseline removal operation and the expected values.
The correction coefficient is 1.1$\pm$0.2.

The following analysis employs the method developed in the ADMX experiment \cite{2022CervanteS3}.

\begin{figure}[htp]
  \centering
  \includegraphics[width=1\columnwidth]{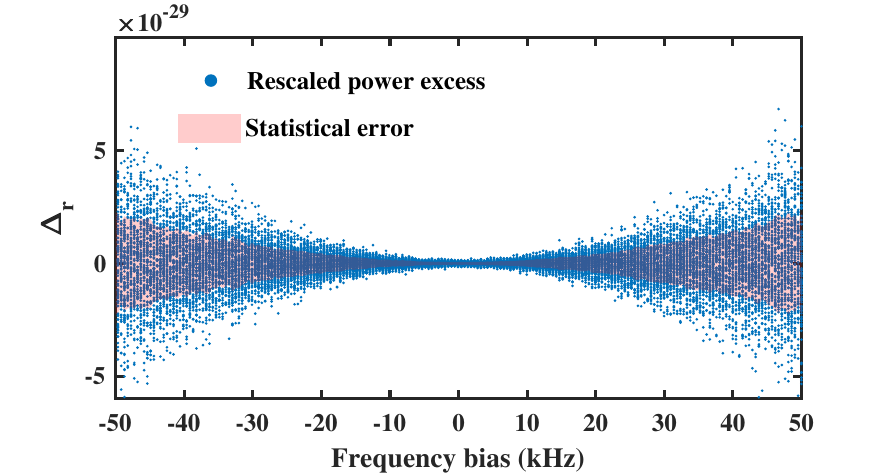}
  \caption{
    The rescaled power excess of all subruns.
    The shallow red ribbon stands for the statistical error.
    }
  \label{FigureS12}
\end{figure}

The normalized power excess indicates the power fluctuation due to the thermal noise.
It is in the unit of $k_{\rm B}T_{\rm n}B$, where $k_{\rm B}$ is the Boltzmann constant, $T_{\rm n}$ is the noise temperature, and $B$ is the frequency bin width of the spectra.
In order to relate the normalized power excess to the real power, each data point is multiplied by $k_{\rm B}T_{\rm n}B$.
Since the power of the expected dark photon signal varies depending on its central frequency, the normalized power excess is rescaled to the dark photon signal with $\chi=1$.
The rescaled power excess
\begin{equation}
  \Delta_{\rm r} = \Delta_{\rm s}\frac{k_{\rm B}T_{\rm n}B}{P_{\rm s}(\chi=1)}
\end{equation}
 is plotted in Fig. \ref{FigureS12}, where $P_{\rm s}(\chi)$ is the expected dark photon signal with a specific value of $\chi$.
As a result, the expected dark photon signal would have the same amplitude regardless of its position on the spectrum.

\begin{figure}[htp]
  \centering
  \includegraphics[width=1\columnwidth]{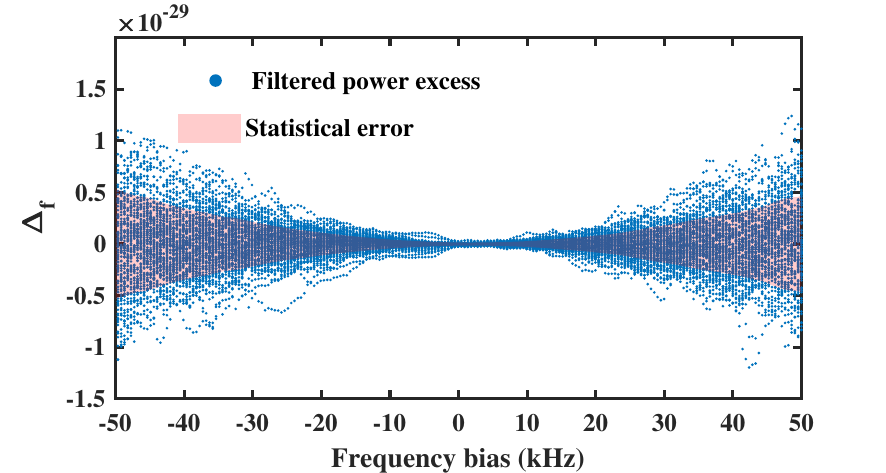}
  \caption{
    The filtered power excess of all subruns.
    The shallow red ribbon stands for the statistical error.
    }
  \label{FigureS13}
\end{figure}

Since the linewidth of the dark photon is 6.52 kHz, the expected dark photon signal would spread over about 15 frequency bins.
As a result, the SNR would be reduced by about 15 times.
Since the amplitude of the expected dark photon signal is already normalized in the previous step, a convolution between the rescaled power excess and the line shape of the dark photon could consolidate the expected dark photon signal into a single bin.
Thus, the SNR would be enhanced.
The convolution operation leads to the spectrum of the filtered power excess $\Delta_{\rm f}$, as Fig. \ref{FigureS13} shows.

\begin{figure}[htp]
  \centering
  \includegraphics[width=1\columnwidth]{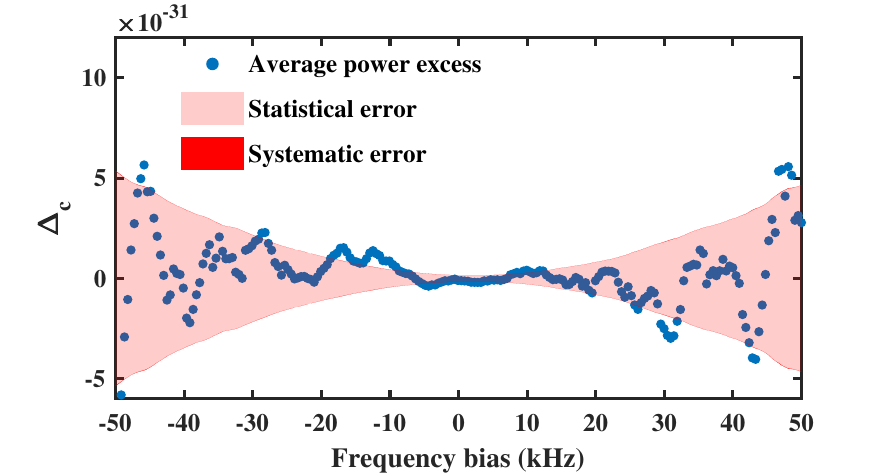}
  \caption{
    The average power excess.
    The value of each data point is the mean of 100 filtered power excess values on the corresponding frequency bin.
    The shallow red ribbon stands for the statistical error.
    The deep red ribbon stands for the systematic error introduced by the uncertainties of $Q_{\rm L}$, $\beta$, $V$, and $T_{\rm n}$.
    }
  \label{FigureS14}
\end{figure}

The 100 filtered power excess spectra obtained individually from each subrun are averaged to further reduce the noise.
The result $\Delta_{\rm c}$ is shown in Fig. \ref{FigureS14}.
The value of each point is averaged from 100 filtered power excess values on the corresponding frequency bin.
The shallow red ribbon refers to the statistical errors $\sigma_{\rm c}$.
The systematic errors are now introduced into the analysis.
The mean value and relative uncertainties of the experimental parameters are shown in Table \ref{Table2}.
Since the uncertainty introduced by the resonant frequency $f_{\rm c}$ is too small, this term is neglected in the analysis.
The systematic errors are combined with the statistical errors as $\sigma'_{c} = \sqrt{\sigma_{c}^{2}+\Delta_{c}^{2}(\sigma_{Q_{\rm L}}^{2}+\sigma_{\beta}^{2}+\sigma_{V}^{2}+\sigma_{T_{\rm n}}^{2})}$, where $\sigma_{Q_{\rm L}}$, $\sigma_{\beta}$, $\sigma_{V}$, $\sigma_{T_{\rm n}}$ are the relative uncertainties of $Q_{\rm L}$, $\beta$, $V$ and $T_{\rm n}$, respectively.
The contribution of the systematic errors to the total errors is shown as the deep red ribbons in Fig. \ref{FigureS14}.
\begin{table}
  \linespread{1.5}
   \caption{Summary of the mean values and the relative systematic uncertainties of experimental parameters.}
   \label{table1}
   \begin{tabular}{l c c}
    \hline
    \hline
   Parameter  &Value & relative uncertainty \\
   \hline
      {$Q'_{\rm L} = Q_{L}Q_{a}/(Q_{L}+Q_{a})$}               & $3.8\times 10^{5}$       & $9.3\times 10^{-4}$\\
      {$\beta$}                                               & 0.4                      & $9.7\times 10^{-4}$\\
      {$f_{\rm c}$}                                           & $6.5$ GHz                & $2.1\times 10^{-9}$\\
      {$V$}                                                   & $1.5\times 10^{-1}$ L    & $5.7\times 10^{-3}$\\
      {$T_{\rm n}$}                                           & $3.0\times 10^{-1}$ K    & $7.4\times 10^{-2}$\\
      \hline
      \hline
    \end{tabular}
  \label{Table2}
\end{table}

\section{Bayesian analysis}

Since no candidate dark photon signal has been discovered, as Fig. 3 (b) in the main text shows, Bayesian analysis is employed to provide constraints of the kinetic mixing $\chi$ of the dark photon.
For a given dark photon signal $P_{\rm s}$, the measured power excess $\widetilde{\Delta}_{\rm c}$ would distribute around $P_{\rm s}$ in a Gaussian form:
\begin{equation}
  p(\widetilde{\Delta}_{\rm c}|P_{\rm s}) = \frac{1}{\sqrt{2\pi}\sigma_{\rm c}}\exp[-\frac{(\widetilde{\Delta}_{\rm c}-P_{\rm s})^{2}}{2\sigma_{\rm c}^{2}}].
  \label{p1}
\end{equation}
Note that the dark photon signal $P_{\rm s}$ is in the unit of $P_{\rm s}(\chi=1)$, so it can be replaced by $\chi^{2}$:
\begin{equation}
  p(\widetilde{\Delta}_{\rm c}|\chi^{2}) = \frac{1}{\sqrt{2\pi}\sigma_{\rm c}}\exp[-\frac{(\widetilde{\Delta}_{\rm c}-\chi^{2})^{2}}{2\sigma_{\rm c}^{2}}].
  \label{p1}
\end{equation}
However, we are actually in an opposite condition where the power excess $\Delta_{\rm c}$ is a known parameter while $P_{\rm s}$ is unknown.
Therefore, Bayes' theorem is employed to estimate the distribution of $P_{\rm s}$:
\begin{equation}
  p(\chi^{2}|\Delta_{\rm c}) = p(\widetilde{\Delta}_{\rm c} = \Delta_{\rm c}|\chi^{2})\frac{p(\chi^{2})}{p(\widetilde{\Delta}_{\rm c} = \Delta_{\rm c})},
  \label{p2}
\end{equation}
  where $p(\chi^{2})$ and $p(\widetilde{\Delta}_{\rm c})$ are the probability density function of $\chi^{2}$ and $\widetilde{\Delta}_{\rm c}$ respectively.
The distribution of $\chi^{2}$ should be uniform on the interval $[0,\chi^{2}_{\rm Q}]$, because $\chi^{2}<0$ is impossible while $\chi^{2}>\chi^{2}_{\rm Q}=3\times 10^{-24}$ have been excluded by the QUALIPHIDE experiment \cite{2023Ramanathan}.
Since $\widetilde{\Delta}_{\rm c}$ has a fixed value from the measurement, the distribution of $\widetilde{\Delta}_{\rm c}$ should be a delta function, $p(\widetilde{\Delta}_{\rm c}) = \delta(\widetilde{\Delta}_{\rm c}-\Delta_{\rm c})$.
Here we replace the delta function with the limit of a finite function,
\begin{equation}
  \begin{aligned}
    \delta(x) &= \lim_{\kappa\to\infty} \xi(x;\kappa),\\
    \xi(x;\kappa)  &= \left\{
                    \begin{array}{lr}
                      1/\kappa, &0<x<\kappa\\
                      0, &\rm otherwise\\
                    \end{array}
                  \right.
  \end{aligned}
  \label{delta}
\end{equation}
  in order to avoid the problem of divergence.

Consequently, the conditional distribution of $\chi^{2}$ can be simplified to
\begin{equation}
  \begin{aligned}
    p(\chi^{2}|\Delta_{\rm c}) &= \frac{p(\widetilde{\Delta}_{\rm c} = \Delta_{\rm c}|\chi^{2})p(\chi^{2})/p(\widetilde{\Delta}_{\rm c} = \Delta_{\rm c})}{\int p(\widetilde{\Delta}_{\rm c} = \Delta_{\rm c}|\chi^{2})p(\chi^{2})/p(\widetilde{\Delta}_{\rm c} = \Delta_{\rm c})\,d\chi^{2}}\\
                               &= \frac{p(\Delta_{\rm c}|\chi^{2})}{\int_{0}^{\chi^{2}_{\rm Q}} p(\Delta_{\rm c}|\chi^{2})\,d\chi^{2}}\\
  \end{aligned}
  \label{p3}
\end{equation}
Finally, by solving the equation
\begin{equation}
  \begin{aligned}
    \int_{0}^{\chi_{90\%}^2} p(\chi^{2}|\Delta_{\rm c}) d\chi^{2} = 90\%
  \end{aligned}
  \label{p90}
\end{equation}
  for each frequency bin, the constraints of the kinetic mixing with a confidence level of $90\%$ are obtained, as Fig. 4 (a) in the main text shows.

\end{document}